\documentclass[10pt]{article}

\usepackage{microtype}
\usepackage{graphicx}
\usepackage{subfigure}
\usepackage{booktabs} 

\usepackage{hyperref}

\usepackage[accepted]{icml2023}

\usepackage{amsmath}
\usepackage{amssymb}
\usepackage{mathtools}
\usepackage{amsthm}

\usepackage{colortbl}
\usepackage{multirow}
\usepackage{multicol}
\usepackage{array}

\usepackage{booktabs} 
\usepackage{multirow} 

\usepackage[capitalize,noabbrev]{cleveref}

\usepackage{chemfig}
\setchemfig{atom sep=2em} 
\theoremstyle{plain}

\theoremstyle{definition}

\theoremstyle{remark}

\usepackage[textsize=tiny]{todonotes}

\icmltitlerunning{A Review on Fragment-based De Novo 2D Molecule Generation}

\begin{document}

\twocolumn[
\icmltitle{A Review on Fragment-based De Novo 2D Molecule Generation}



\icmlsetsymbol{equal}{*}

\begin{icmlauthorlist}
\icmlauthor{Sergei Voloboev}{yyy}

\end{icmlauthorlist}

\icmlaffiliation{yyy}{School of Computation, Information and Technology, TUM, Munich, Germany}

\icmlcorrespondingauthor{Sergei Voloboev}{voloboeff201087@gmail.com}

\icmlkeywords{Machine Learning, ICML}

\vskip 0.3in
]



\printAffiliationsAndNotice{} 

\begin{abstract}
In the field of computational molecule generation, an essential task in the discovery of new chemical compounds, fragment-based deep generative models are a leading approach, consistently achieving state-of-the-art results in molecular design benchmarks as of 2023. We present a detailed comparative assessment of their architectures, highlighting their unique approaches to molecular fragmentation and generative modeling. This review also includes comparisons of output quality, generation speed, and the current limitations of specific models. We also highlight promising avenues for future research that could bridge fragment-based models to real-world applications.
\end{abstract}

\section{Introduction}
The search for new molecules is a major challenge, central to the development of new materials and the synthesis of new drugs. The theoretical compound space is staggeringly large, estimated between $10^{23}$ and $10^{60}$ potential compounds \cite{Bilodeau2022GenerativeMF}. Efficiently navigating this immense space is critical, and deep generative models have emerged as a key solution. Early generative models focused on 1D string representations of molecules, specifically SMILES (Simplified Molecular Input Line Entry System) \cite{Weininger1988SMILESAC}. Pioneering work in this area \cite{GmezBombarelli2016AutomaticCD, Kusner2017GrammarVA} has highlighted the potential of these models. However, SMILES has limitations in capturing structural similarities - structurally similar molecules can have significantly different SMILES strings. In addition, ensuring chemical validity in 1D models is complex. These challenges have led to a shift toward 2D approaches to molecule generation, which represent molecules as graphs, with atoms and bonds represented as nodes and edges, respectively. While constructing molecules atom by atom seems straightforward, intermediate structures often exhibit unusual chemical fragments, such as incomplete aromatic rings \cite{Maziarz2021LearningTE}. Fragment-based methods address these problems by using common molecular fragments as the primary units for generation, ensuring chemically valid structures at each step. 

While 2D fragment-based models have achieved state-of-the-art results in generative benchmarks, their graph representation does not represent the exact spatial structure, which can be a problem for tasks that require this information, such as target protein binding. In recent years, 3D generative models that address this issue have gained attention in the research community. However, the latest 3D methods \cite{xu2023geometric, hoogeboom2022equivariant} do not yet match the generative capabilities of 2D fragment-based models on generative benchmarks.

This paper provides an overview of existing fragment-based architectures for 2D molecular generative modeling. \Cref{sec:theory} introduces the conceptual framework of the topic. \Cref{sec:methods} outlines our categorization of the reviewed methods. Performance aspects and important details are discussed in \autoref{sec:discussion}. Finally, \autoref{sec:future} summarizes the papers and explores possible directions for future research.

\begin{table*}[hbt!]
\centering
\caption{Classification of fragment-based generative models for de novo 2D molecule generation.}
\begin{tabular}{@{} cccccc @{}}
\toprule
Generative & & Presence of & Generation &  & Vocabulary \\
Architecture & Method & Hypergraph & Approach & Decomposition & Construction \\
\midrule
\multirow{6}{*}{VAE} & JT-VAE & $\checkmark$ & Sequential & Chemically inspired & All fragments \\
 & HierVAE & $\checkmark$ & Sequential & Chemically inspired & All fragments \\
 & MAGNet & $\checkmark$ & All at once & Chemically inspired & All fragments \\
 & MoLeR & & Sequential & Chemically inspired & Common fragments (motifs) \\
 & MicaM & & Sequential & Data-driven & Data-driven \\
 & PS-VAE & & All at once & Data-driven & Data-driven\\
\midrule
MCMC sampling & MARS & & Sequential & Chemically inspired & Common fragments (motifs) \\
\bottomrule
\end{tabular}
\label{tab:classification}
\end{table*}

\section{Theory}
\label{sec:theory}

    In the field of computational chemistry, \textit{de novo design} holds an important place, particularly for its role in automating the creation of new compounds \cite{Schneider2005ComputerbasedDN}. Central to this process is the concept of a \textit{molecular graph} \( G \), defined as the tuple \( \{V, E\} \). Here, \( V \) denotes the multiset of atoms of size \( n = |V| \), and \( E \) represents the bond tensor, formatted as \( \{0, 1\}^{n \times n \times 4} \). This tensor uses one-hot encoding to indicate either the absence of a bond or the presence of single, double, or triple bonds.

    The goal of 2D generative molecule modeling is to \textit{learn the distribution} \( P(G) \) over a set of graphs \( \mathcal{G} = \{G_i\} \) and then sample new graphs \( G_{\text{new}} \) from this distribution \cite{sommer2023the}.

While a \textit{fragment} is defined as any connected subgraph of a molecule, a \textit{motif} refers specifically to a frequently occurring fragment. Motifs are similar to functional groups in chemistry, such as amine (\( -NH_2 \)) groups.


\textit{Fragment-based methods} treat molecules as multisets of fragments \( \mathcal{F} = [F_{1}, \ldots, F_{n}] \). This approach not only guarantees the chemical validity of the generated compounds \cite{hu2023deep}, but also improves interpretability and control over the generation process \cite{Jin2020MultiObjectiveMG} as experts can customize the fragment vocabulary to prevent the generation of unwanted structures.

\section{Methods}
\label{sec:methods}

Almost all fragment-based distributional learning models reviewed at this time are based on Variational Autoencoders (VAEs)~\cite{Kingma2013AutoEncodingVB}, and we consider the following VAE-based algorithms: JT-VAE~\cite{Jin2018JunctionTV}, HierVAE~\cite{jin2020hierarchical}, MoLeR~\cite{Maziarz2021LearningTE}, PS-VAE~\cite{kong2022molecule}, MiCaM~\cite{geng2023novo}, and MAGNet~\cite{hetzel2023magnet}. These methods share a common training pipeline: fragmenting a molecule, encoding it into a latent space, and decoding it back to generate the molecule. We review the common strategies used by these models to perform these steps.

As a notable exception, MARS~\cite{Xie2021MARSMM} differs from the VAE approach by using Markov Chain Monte Carlo (MCMC) sampling. Originally aimed at goal-directed generation, it also shows strong performance in distribution learning benchmarks~\cite{kong2022molecule}. Unlike VAE, MARS lacks a molecule encoding stage and does not use latent representations \( z \). It starts with a
base structure of two carbon atoms and samples molecular
editing actions based on Maximum Likelihood Estimation
(MLE), conditional on the current state of \( \mathcal{G}^{'}\), so its generation process is similar to the sequential generation in VAEs, as described in \autoref{sec:decoding}.


While atom-by-atom methods have used Generative Adversarial Networks (GANs)~\cite{DeCao2018MolGANAI} and Normalizing Flow~\cite{Shi2020GraphAFAF}, fragment-based approaches using these generative paradigms are absent from the current literature. 

The axes of classification of fragment-based generative algorithms are discussed in the \autoref{sec:decomposition} and the \autoref{sec:distr_learning}. The classification of the reviewed algorithms is presented in \cref{tab:classification}.

\subsection{Fragmentation scheme}
\label{sec:decomposition}
 The decomposition of a molecular graph \( G \) into a collection of fragments \( \mathcal{F} \) is typically approached through two different paradigms - \textit{chemically inspired} and \textit{data-driven} approaches. The differences between them are reflected in corresponding fragmentation schemes that specify how to decompose molecules into fragments and create vocabularies of fragments \( V \) \cite{sommer2023the}.
\subsubsection{Decomposition}
The decomposition phase involves the partitioning of a molecular graph \( G \) into a multiset of fragments \( \mathcal{F}\). \textit{Chemically inspired} techniques use hand-crafted heuristics to achieve fragmentation. Common to these algorithms is a design principle that avoids breaking bonds within structurally significant units, such as aromatic rings. In contrast, \textit{data-driven} methods segment molecules into fragments based on a previously compiled motif vocabulary \( V \).

\subsubsection{Vocabulary Construction}
Some \textit{chemically inspired} strategies include all fragments derived from the training set in the fragment vocabulary \( V \). Others, such as MoLeR, limit the size of the vocabulary to \( |V| = k \), where \(k\) is some predefined limit on the maximum number of fragments in the vocabulary, including only the most common fragments that can be considered motifs. Since not all the fragments are included, these methods always use atoms as additional generation units. 

In contrast, \textit{data-driven} models construct their vocabularies through a bottom-up mechanism similar to Byte Pair Encoding (BPE) \cite{Gage1994ANA}. Starting with vocabularies composed of single atoms, these models iteratively merge frequent adjacent fragment pairs to create increasingly complex motifs. While the data-driven approach to vocabulary construction can be viewed as extracting motifs from the dataset (similar to the inclusion of the most frequent fragments), we separate them into distinct categories in our classification.

    
    

\subsection{Distributional Learning}
\label{sec:distr_learning}
\subsubsection{Molecule Encoding}
Molecular encoding is performed in a unified manner in the VAE methods under study. The process starts by computing an embedding for a molecule \( G_i \), formulated as \( h_{G_i} = \text{GNN}(G_i) \). Next, a latent vector \( z \) is sampled from a posterior distribution \( q(z|G_i) \), characterized as a normal distribution \( \mathcal{N}(\mu(h_{G_i}), \Sigma(h_{G_i})) \)

\subsubsection{Generating a molecule}
 \label{sec:decoding}
To generate a molecule from its latent representation \( z \) in VAE models, we categorize the existing methods into two distinct groups, drawing inspiration from previous studies \cite{yang2022molecule}: \textit{sequential approaches} and \textit{all at once approaches}.

\textit{Sequential approaches} select each new fragment \( F_{i} \), conditioned on the evolving molecular structure \( \mathcal{G}^{'} \). Upon sampling \( F_{i} \) from \( P(F_{i} | \mathcal{G}^{'}, z) \), it is immediately integrated into \( \mathcal{G}^{'} \), forming bonds between atoms in \( F_{i} \) and \( \mathcal{G}^{'} \).
Teacher forcing is commonly used in these methods to guide subsequent predictions \cite{Jin2018JunctionTV}. 

Similarly, MARS sequentially generates and integrates fragments into \( \mathcal{G}^{'} \), although it operates without guidance from a latent representation \( z \). A feature of MARS that distinguishes it from other sequential approaches is its ability to delete existing bonds in \( \mathcal{G}^{'} \).

Conversely, \textit{all at once approaches} first generate the entire multiset of fragments \( \mathcal{F} = [F_{1}, \ldots, F_{n}] \) in an autoregressive manner using \( P(F_{i} | F_{<i}, z) \). Once \( \mathcal{F} \) is generated, these methods simultaneously estimate \( P(e_{uv} | \mathcal{F}, z) \) for all possible intra-fragment connections.

\subsubsection{Presence of hypergraph at decoding stage}
While the previously described models directly generate bonds between atoms from different fragments, certain approaches first construct a hypergraph \(\mathcal{H}_G = (\mathcal{F}, \mathcal{E})\), where \( \mathcal{E}\) denotes edges between fragments, with each edge \(e_{ij}\) identifying the connection between fragments \(F_i\) and \(F_j\) and having the set of shared atom types as a feature. Once the fragment-level edges are determined, these models then select specific atoms from \(F_i\) and \(F_j\) to form bonds. 

The methodology for constructing such a hypergraph varies, again involving either a \textit{sequential} or an \textit{all at once} strategy.

JT-VAE and HierVAE decompose the molecule into a tree-like structure that is decoded \textit{sequentially} in a depth-first order.

In contrast, MAGNet first generates a multiset of shapes (a shape is a collection of fragments with an identical adjacency matrix) and then infers the connectivity of these shapes in an \textit{all at once} fashion.

\section{Discussion}
\label{sec:discussion}


\textbf{Performance Comparison and Datasets.}  Evaluating the quality of the generated samples of a molecule generation method is the challenge without a single solution. Early studies typically measured \textit{validity}, \textit{uniqueness}, and \textit{novelty}. Validity refers to the percentage of molecules that pass the valence check, uniqueness indicates the percentage of non-duplicate molecules, and novelty represents the percentage of molecules that are not found in the training set. 
However, in recent models, achieving 100\% validity and nearly 100\% uniqueness and novelty has become common, as seen in studies such as Maziarz et al. \yrcite{Maziarz2021LearningTE}, Hetzel et al. \yrcite{hetzel2023magnet}, and Geng et al. \yrcite{geng2023novo}, and these metrics are often not reported separately. 

As a result, more informative metrics emerged, including the \textit{KL divergence} between the distributions of physicochemical properties of the generated molecules and those in the training set, and the \textit{Fréchet ChemNet Distance} (FCD), which assesses the similarity between the hidden representations of the generated molecules and those in the training set, as determined by ChemNet \cite{preuer2018frechet}.

When it comes to datasets,
QM9 \cite{ruddigkeit2012enumeration}, which contains all possible organic molecules with up to 9 heavy atoms (excluding hydrogen), 
has been used extensively in molecule generation research due to its status as one of the earliest datasets. However, it is not very suitable for learning the distribution of drug-like molecules, since it contains many complex compounds that have not yet been synthesized \cite{brown2019guacamol}. In addition, it is relatively small compared to new datasets, having only 134,000 molecules.

Therefore, most recent papers find it insufficient to report results based on it alone. Another popular dataset is ZINC250K \cite{Kusner2017GrammarVA}, which is a subset of 250,000 molecules from the ZINC database \cite{irwin2020zinc20} that contains 3D structural information of drug-like molecules and molecular physical properties.

The largest current dataset for de novo molecular design is GuacaMol \cite{brown2019guacamol}, which contains 1,591,378 drug-like molecules extracted from the ChEMBL 24 database \cite{mendez2019chembl}.
It is important to note that the same name "GuacaMol" is used in the literature for the distributional learning benchmark, the goal-directed generation benchmark, and the dataset presented in the Brown et al. paper \cite{brown2019guacamol}, which causes some confusion for an inexperienced reader.

During the analysis of the literature, we found several discrepancies in the results, so we don't present the exact numbers in this overview. We can only speculate that MiCaM, MoLeR, or MAGNet are potential state-of-the-art (SOTA) candidates, and a more comprehensive and holistic evaluation would help select the best algorithm.

\textbf{Runtime and Time Complexity.} The generation speed of the reviewed models is critical for practical purposes, especially since these models often generate thousands of new molecules in downstream tasks. However, comprehensive runtime benchmarks are lacking in the literature. We can only note that methods like PS-VAE and MoLeR outperform earlier techniques like JT-VAE in training and inference speed by orders of magnitude \cite{kong2022molecule, Maziarz2021LearningTE}.

Regarding time complexity, these methods typically exhibit a worst-case complexity of \( \mathcal{O}(n^2) \) for \( n \) atoms. However, the practical average-case complexity is often lower due to restrictions on inter-fragment edges and constraints such as molecular symmetries \cite{Maziarz2021LearningTE} or connection-aware motifs \cite{geng2023novo}. The relationship between generation and fragmentation also has a significant impact on generation runtime.


\textbf{Goal-Directed Generation and Lead Optimization.} 
A typical downstream task for which the reviewed approaches are used is the generation of molecules with specific properties, either de novo or from a lead structure.
While many reviewed methods can adapt to these tasks through latent space optimization techniques such as Molecular Swarm Optimization (MSO) \cite{winter2019efficient}, our review focuses only on the distributional learning aspect. In the area of goal-directed optimization, there are several alternative methods that do not depend on distributional learning, including RL-based approaches~\cite{Yang2021HitAL, Jin2020MultiObjectiveMG} and genetic algorithms~\cite{nigam2019augmenting}.

\textbf{Synthesizability.}
An often neglected but critical factor for real-world downstream tasks is the synthesizability of the generated molecules. The methods reviewed generally either neglect this aspect or rely on simplistic metrics such as Synthetic Accessibility (SA) scores~\cite{Ertl2009EstimationOS}. However, these scores can be misleading as complex models can easily circumvent their limitations~\cite{Bilodeau2022GenerativeMF}.

\textbf{Motif Vocabulary Size and Fragment Design Considerations.} 
Another frequently overlooked aspect is the size of the motif vocabularies of the models. A larger vocabulary leads to a better fit to the training set~\cite{Maziarz2021LearningTE}, so it is crucial to use a similar number of fragments in the benchmarked methods for a fair comparison.

For example, one SOTA candidate, MiCaM, uses connection-aware motifs with a vocabulary size of 4096, which is significantly larger than the vocabulary sizes of the methods it was compared to. This larger vocabulary likely contributes to its high performance.

Another SOTA contender, MoLeR, explicitly integrates atoms as generation units. This approach not only appears to improve the fit to the target distribution, but also provides greater flexibility for fine-tuning with novel molecules. In contrast, purely motif-based models such as JT-VAE face challenges in adapting to new fragment types, requiring vocabulary rebuilding and model retraining.


\textbf{Fragment Generation Order.} 
All of the methods discussed adhere to a specific fragment generation order (uniform random, BFS, canonical, etc.), which contradicts the multiset nature of fragments in a molecule. Analysis of different generation orders within the MoLeR architecture revealed that random order underperforms fixed order \cite{Maziarz2021LearningTE}. This suggests that sequential fragment decoders, such as in MoLeR, are sensitive to biases introduced by fragment order. 
Current "all at once" generation methods use either a random \cite{kong2022molecule},  or a sorted by shape order \cite{hetzel2023magnet} for fragment generation. Investigating the performance of approaches that utilize equivariant set generation techniques \cite{zhang2019deep, vignac2021top} could improve the quality and size of the generated samples \cite{Faez2020DeepGG}.





\section{Conclusion and Future Research}
\label{sec:future}
In this paper, we have provided an overview of fragment-based models for 2D molecular generation, focusing on their fragmentation and generative strategies, including both sequential and 'all at once' approaches. Various performance considerations have been discussed and existing challenges in the field have been addressed. Looking forward, we identify these research directions as promising for the field:

 \textbf{Transparency and Uncertainty Quantification. } Understanding how models decide on specific fragment is crucial for critical areas such as drug design \cite{Deng2021ArtificialII}. Integrating uncertainty quantification into models can identify reliable molecules with promising properties, thereby streamlining the drug design process \cite{hu2023deep}. 
    
\textbf{Focus on Realistic Properties. }
    While many studies focus on computational metrics such as logP or QED11, there is a need for experimental validation of lead molecules. Future research could shift towards the generation of molecules optimized for properties that are realistically testable and applicable \cite{Bilodeau2022GenerativeMF}.
    
\textbf{Advancing 3D Molecule Generation.} The transition to 3D molecular representations is critical for structure-specific tasks. The application of fragment-based methods to 3D molecule generation provides an opportunity to address validity challenges in this area. Despite initial studies~\cite{qiang2023coarse}, further exploration in this area is needed.



\bibliography{main}
\bibliographystyle{icml2023}



\end{document}